\documentclass[letterpaper,12pt]{article}   
\usepackage{osajnl2} 
\usepackage[draft]{hyperref} 
\usepackage{amsmath}
\usepackage{amssymb}
\begin{document}

\title{Nonlinear interaction of two trapped-mode resonances in a~bilayer fish-scale metamaterial}

\author{Vladimir R. Tuz,$^{1,2,*}$ Denis V. Novitsky$^{3}$, Pavel L. Mladyonov$^{1}$, \\ Sergey L. Prosvirnin$^{1,2}$, Andrey V. Novitsky$^{4}$ }
\address{$^1$Institute of Radio Astronomy of National Academy of
Sciences of Ukraine, \\ 4, Krasnoznamennaya st., Kharkiv 61002,
Ukraine}
\address{$^2$School of Radio Physics, Karazin Kharkiv
National University, \\ 4, Svobody Square, Kharkiv 61077, Ukraine}
\address{$^3$B.~I.~Stepanov Institute of Physics of National Academy
of Sciences of Belarus, \\ 68, Nezavisimosti Avenue, Minsk 220072,
Belarus}
\address{$^4$Department of Theoretical Physics, Belarusian State
University, \\ 4, Nezavisimosti Avenue, Minsk 220050, Belarus}
\address{$^*$Corresponding author: tvr@rian.kharkov.ua}

\begin{abstract} We report on a bistable light transmission through a bilayer
``fish-scale" (meander-line) metamaterial. It is demonstrated that
an all-optical switching may be achieved nearly the frequency of the
high-quality-factor Fano-shaped trapped-mode resonance excitation.
The nonlinear interaction of two closely spaced trapped-mode
resonances in the bilayer structure composed with a Kerr-type
nonlinear dielectric slab is analyzed in both frequency and time
domains. It is demonstrated that these two resonances react
differently on the applied intense light which leads to destination
of a multistable transmission.
\end{abstract}


\maketitle 

\section{Introduction}
\label{sec:intro}

During recent years research efforts on both fundamental physics and
applications of nonlinear optics demonstrate several major trends
\cite{Kivshar_08}. First, the basic concepts of nonlinear optics
penetrated into new areas of material science by exploring novel
nonlinear materials and nonlinear propagation of light in engineered
structures. Second, there are a number of novel concepts related to
tunable nonlinear response, and engineered and enhanced
nonlinearities, which should be explored more extensively for
developing advanced optical tunable nonlinear devices. Moreover, in
such structures many of the effects well studied in nonlinear optics
can be enhanced by the cavity effects, heterostructures, or
Fano-type resonances, that enable much stronger nonlinear response
at lower powers. Metamaterials based on planar resonant patterns are
expected to become the key building blocks for such
nonlinearity-enhanced effects and devices.

It is believed that the fabrication of metal-dielectric composite
structures where the dielectric layers possess strong nonlinear
response can allow the realization of tunable nonlinear
metamaterials. The important area where the role of such nonlinear
structures is expected to become more and more pronounced is the
creation of all-optical circuitry for computing, information
processing, and networking which is expected to overcome the speed
limitations of electronics-based systems.

Typically metamaterials consist of arrays of magnetically or
electrically polarizable elements. In many common configurations,
the functional metamaterial inclusions are planar metallic patterns
composed of \textit{symmetrical} split-ring resonators (SRRs), on
which currents are induced that flow in response to incident
electromagnetic fields. A distinct property of metamaterials is that
the capacitive regions of SRRs strongly confine and enhance the
local electric field. The capacitive region of a metamaterial
element thus provides a natural entry point at which some lumped
frequency-dispersive, active, tunable or nonlinear materials can be
introduced, providing a mechanism for coupling the metamaterial
geometric resonance to other fundamental material properties
\cite{soukoulis_OptExpress_2008,shadrivov_APL2008,poutrina_NewJPhys_2010}.
The hybridization of metamaterials with semiconductors,
ferromagnetic/ferrimagnetic materials and other externally tunable
materials has already expanded the realm of metamaterials from the
passive, linear media initially considered.

While the wave propagation can be controlled by using metamaterials
with a unit cell consisting of a single resonant element
(symmetrical split-ring resonator), other characteristics can also
be controlled by using metamaterials composed of coupled resonators.
Such metamaterials sometimes are also referred to EIT-like
metamaterials \cite{papasimakis_ApplPhysLett_2009,Kitano2012}.
Typically they consist of identical subwavelength metallic
inclusions structured in the form of \textit{asymmetrical}
split-rings (ASRs)
\cite{prosvirnin_Chapter_2003,fedotov_PhysRevLett_2007}, split
squares \cite{khardikov_JOpt_2010} or their complementary patterns
\cite{samson_ApplPhysLett_2010}. These elements are arranged
periodically and placed on a thin dielectric substrate. It is
revealed that since these particles possess specific structural
asymmetry, in a certain frequency band, the antiphase current
oscillations with almost the same amplitudes appear on the particles
parts (arcs). The scattered electromagnetic far-field produced by
such current oscillations is very weak, which drastically reduces
its coupling to free space and therefore radiation losses. Indeed,
both the electric and magnetic dipole radiations of currents
oscillating in the arcs of particles are canceled. As a consequence,
the strength of the induced current can reach very high value and
therefore ensure a high-$Q$ resonant optical response. Such a
resonant regime is referred to so-called ``trapped-modes", since
this term is traditionally used in describing electromagnetic modes
which are weakly coupled to free space.

The features of nonlinear trapped-mode resonances were previously
studied theoretically in planar metamaterials composed of
asymmetrical split-rings \cite{tuz_PhysRevB_2010} and two concentric
rings \cite{tuz_EurPhysJApplPhys_2011, tuz_JOpt_2012,
dmitriev_AdvElectromagn_2012}. In both configurations the substrate
which carries the metallic pattern is considered as a nonlinear
dielectric. The effect of nonlinearity appears as the formation of a
closed loop of bistable transmission within the frequency of
trapped-mode resonance. Since the nonlinear response of these
metamaterials operating in the trapped-mode regime is extremely
sensitive to the dielectric properties of the substrate it allows
one to control switching operation effectively.

Another type of structures, which bears Fano-shaped trapped-mode
resonances and is very promising for applications, is planar
metamaterials consisted of equidistant array of continuous meander
or zigzag metallic strips placed on a thin dielectric substrate
(fish-scale structures \cite{fedotov_PhysRevE_2005,
zhang_OptLett_2012, rance_PhysRevB_2012}). In the past such
structures have been investigated both theoretically
\cite{fedotov_PhysRevE_2005, mladyonov_PhysicsAstronomy_2010,
prosvirnin_JElectromagWavesAppl_2002,
kawakatsu_JElectromagWavesAppl_2010} and experimentally
\cite{fedotov_PhysRevE_2005}. It is revealed, when the wave is
polarized orthogonally to the strips, the fish-scale structure is
transparent across a wide spectral range apart from an isolated
wavelength. In contrary, when the wave is polarized parallel to the
strips, it becomes a good broad-band reflector apart from an
isolated wavelength where transmission is high. The same array
combined with a homogeneous metallic mirror is also a good reflector
which preserves a phase of the reflected wave with respect to the
incident wave at this particular wavelength
\cite{fedotov_PhysRevE_2005, Mladenov_2003}. The latter phenomenon
is known as a ``magnetic mirror" \cite{Sievenpiper_1999}. Finally,
the structure also acts as a local field concentrator and a resonant
multifold ``amplifier" of losses in the constitutive dielectric.

Among others, a trapped-mode resonance can be excited in a
fish-scale structure if the incident field is polarized along the
strips and when the form of these strips is slightly different from
the straight line. In such a system the less the form of strips is
different from the straight line, the greater is the quality factor
of the trapped-mode resonance. But, unfortunately, as the line
curvature decreases, the resonant frequency shifts to the frequency
where the Rayleigh anomaly occurs which diminishes the degree of the
field localization within the system.

Nevertheless, in a bilayer configuration of the metamaterial, in
which two equal gratings are adjusted to each side of a dielectric
slab, it is possible to excite the high-$Q$ trapped-mode resonance
far from the Rayleigh anomaly
\cite{mladyonov_PhysicsAstronomy_2010}. This resonance appears due
to the oscillations of currents which flow in opposite directions
relative to the adjacent gratings. In this case the electromagnetic
field is strongly localized in the area between the adjacent
gratings. Furthermore, in the bilayer structure, besides the
trapped-mode resonance excited by gratings, some interference
resonances appear in the similar manner as in traditional systems
with gratings of straight lines. Therefore the bilayer fish-scale
structure allows one to obtain different resonant features. The most
important thing is that in the bilayer fish-scale structure the
trapped-mode resonance has quality factor which is sufficiently
greater than that ones attainable in the convenient interference
resonances.

We should also note that the trapped-mode resonances can be excited
in a very thin bilayer structure which is important for practical
applications. We argue that the bilayer configuration is of a great
interest in the case when the substrate is made of a field intensity
dependent material (e.g. a Kerr-type medium) because the strong
field localization between the gratings can significantly enhance
the nonlinear effects. It leads to some particular manifestation of
effects of bistability and multistability nearly the frequencies of
Fano-shaped trapped-mode resonances which is presented in this
paper.

\section{Problem formulation}\label{sec:formulation}
\subsection{Bilayer fish-scale metamaterial} \label{sec:structure}

\begin{figure}[htb]
\centerline{\includegraphics[width=10.0cm]{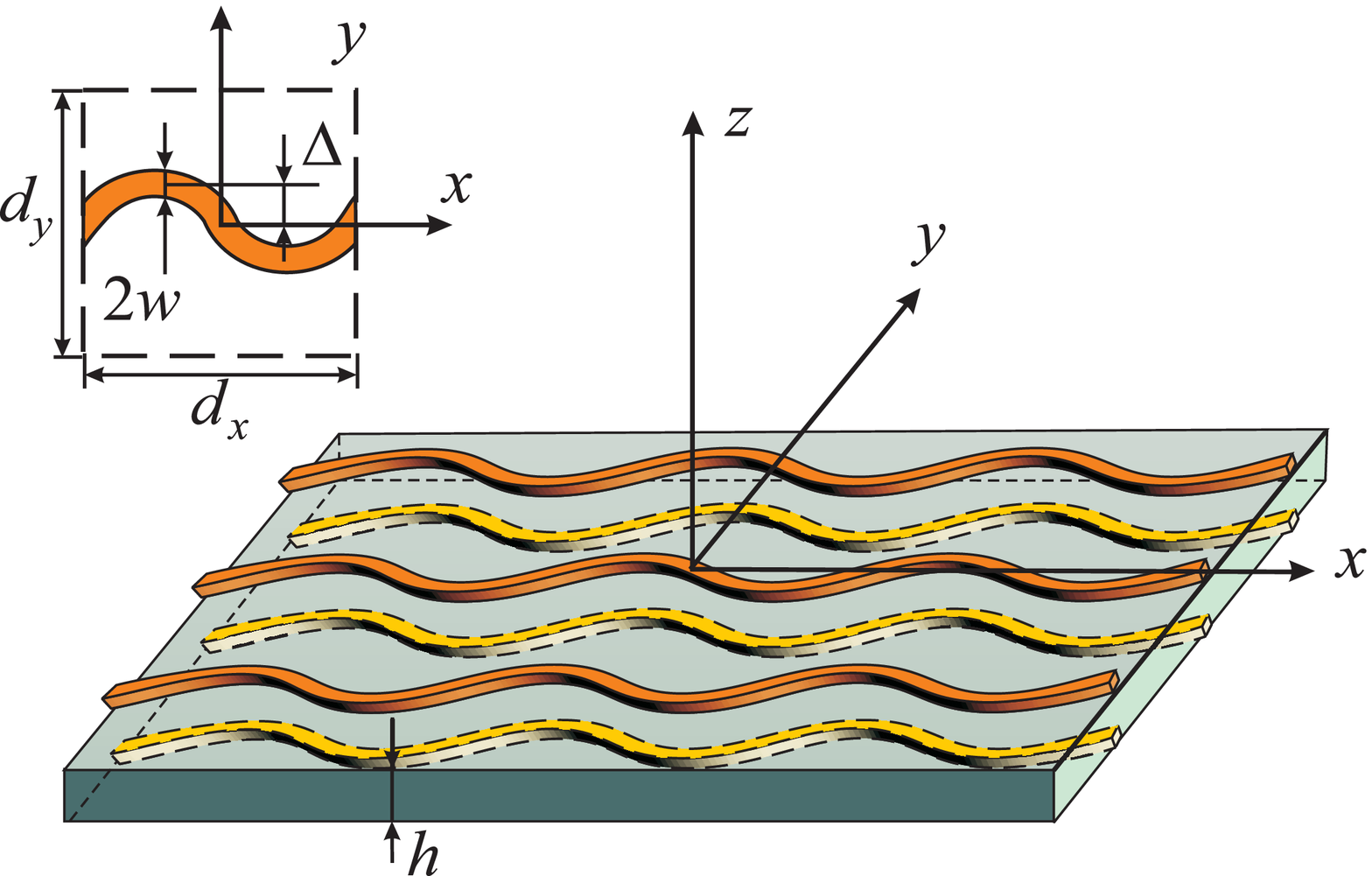}} \caption{}
\label{fig:fig1}
\end{figure}

A studied structure consists of two gratings of planar perfectly
conducting infinite wavy-line strips placed on the both sides of a
dielectric slab with thickness $h$ and permittivity $\varepsilon$
(see Fig.~\ref{fig:fig1}). The elementary translation (unit) cell of
the structure under study is a square with sides $d=d_x=d_y$. The
total length of the strip within the elementary translation cell is
$S$. Suppose that the thickness $h$ and size $d$ are less then the
wavelength $\lambda$ of the incident electromagnetic radiation
($h\ll\lambda$, $d<\lambda$). The width of the metal strips and
their deviation from the straight line are $2w$ and $\Delta$,
respectively.

Assume that the incident field is a plane monochromatic wave
\begin{equation}\label{eq:incident}
\vec E^{inc}=\vec p A_0\exp(-i\vec k^{inc}\vec r),
\end{equation}
where $\vec p$ is the unit vector which defines the polarization of
the incident wave, $|p| = 1$; $A_0$ and $\vec k^{inc}=\vec e_x
k_x^{inc}+\vec e_y k_y^{inc}+\vec e_z k_z^{inc}$ are the magnitude
and  wavevector of the incident wave,  respectively; $\vec e_x$,
$\vec e_y$, $\vec e_z$ are orts of the coordinate system;
$k^{inc}=k=\omega\sqrt{\varepsilon_0\mu_0}$. The time dependence is
supposed to be $\exp(i\omega t)$.

In the frequency domain a solution of the diffraction problem on the
bilayer structure is derived using the method of moments. It is
formulated similarly to that one obtained for a single-layer
metamaterial's configuration
\cite{prosvirnin_JElectromagWavesAppl_2002,
mladyonov_PhysicsAstronomy_2010}. Farther the solution of the linear
problem is used to construct a solution of the nonlinear problem.

We shall seek for the field through the entire space in the form of
a superposition of the field without gratings $\vec E^d$ and the
field scattered by gratings $\vec E^s$
\begin{equation}\label{eq:fieldreftr}
\vec E=\vec E^{d} + \vec E^s.
\end{equation}
Here $\vec E^{d} = \vec E^{inc}+ \mathbb {\vec R}\exp(-i\vec k^s
\vec r)$ and $\vec E^{d} = \vec E^{inc}+ \mathbb {\vec T}
\exp(-i\vec k^{inc} \vec r)$ are the electric field strength in the
areas above ($z \ge 0$) and below ($z \le -h$) the structure,
respectively; $\vec k^s=\{k_x^{inc}, k_y^{inc}, -k_z^{inc}\}$ is the
wavevector of the reflected wave, and vectors $\mathbb {\vec R}$ and
$\mathbb {\vec T}$ are defined using corresponding conditions for
dielectric boundaries.

Further we restrict ourself with a case of normal incidence of
$x$-polarized wave ($k_x^{inc}= k_y^{inc}=0$), i.e. $\vec E^{inc}$
is directed along the strips ($\vec p = \vec e_x$, $H^{inc}_x=0$,
$E^{inc}_x \ne 0$). For such a polarization, the reflection and
transmission coefficients for a dielectric slab without gratings are
$\mathbb R=(Z^2-1)(1-q)[(Z+1)^2-q(Z-1)^2]^{-1}$, $\mathbb
T=4Zq[(Z+1)^2-q(Z-1)^2]^{-1}$, where $Z=\sqrt{\mu/\varepsilon}$ and
$q=\exp(-2ikh\sqrt{\varepsilon\mu})$.

The surface current densities on the metal strips in the planes
$z=0$ and $z = -h$ are denoted by $\vec J_1(\vec \rho)$ and $\vec
J_2(\vec\rho)$, where $\vec \rho$ is imposed by the strip midline.
In the framework of the method of moments, the metal pattern is
treated as a perfect conductor. Based on the boundary conditions of
equality to zero of tangential components of the electric field on
the strips we arrive to a system of two integral equations related
to the currents induced on the strips
\begin{equation}\label{eq:systemint}
\begin{aligned}
\frac{1}{4\pi^2}\sum_{m,n=-\infty}^\infty
\int\limits_{-\infty}^{\infty} \int\limits_{-\infty}^{\infty}\left\{
\vec E_{1t}(\vec J_1, \vec J_2) \right\}(\vec \kappa) \exp\left [
i(\vec \kappa - \vec k^{inc})\vec \rho_{mn} - i\vec \kappa \vec
\rho\right ] d\vec\kappa+
\vec E_{1t}^d(\vec \rho) = 0,~~~~z=0,\\
\frac{1}{4\pi^2}\sum_{m,n=-\infty}^\infty
\int\limits_{-\infty}^{\infty} \int\limits_{-\infty}^{\infty}\left\{
\vec E_{2t}(\vec J_1, \vec J_2) \right\}(\vec \kappa) \exp\left [
i(\vec \kappa - \vec k^{inc})\vec \rho_{mn} - i\vec \kappa \vec
\rho\right ] d\vec\kappa+
\vec E_{2t}^d(\vec \rho) = 0,~z = -h,  \\
\end{aligned}
\end{equation}
where the linear operators $\left\{ \vec E_{1t}(\vec J_1, \vec J_2)
\right\}(\vec \kappa)$ and $\left\{ \vec E_{2t}(\vec J_1, \vec J_2)
\right\}(\vec \kappa)$  of the \textit{linear problem} are
associated to the Fourier transforms of the surface current density
and the tangential components of the radiation field; $\vec
\rho_{mn}=m d_x \vec e_x + n d_y \vec e_y$ is the radius-vector
directed from the unit cell located at the center of coordinates (we
shall number it as (0, 0)) to the unit cell located at the position
($m, n$) for the upper and lower gratings, respectively; $\vec
\kappa$ is the wavevector of spatial spectrum of the field.

Using the method of moments, the system of integral equations
(\ref{eq:systemint}) can be reduced to the system of linear
algebraic equations related to the unknown expansion coefficients of
the surface currents densities on the strips of the first and second
gratings. When these expansion coefficients of the surface currents
densities are found (in the same way as it was done for a
single-layer metamaterial
\cite{prosvirnin_JElectromagWavesAppl_2002}), it is easy to
calculate the magnitude of the spatial harmonics of the field in the
form of some plane waves expansion. It can be derived both above and
below the structure
\begin{equation}\label{eq:systemalg}
\begin{aligned}
\vec E^{s} = \sum_{\alpha,\beta=-\infty}^{\infty} \vec
a_{\alpha\beta}\exp\left\{-i\left [\vec \kappa_
{\alpha\beta}\vec \rho + \gamma(\vec \kappa_{\alpha\beta}) z  \right ] \right\} ,~~~~z \ge 0,~~\\
\vec E^{s} = \sum_{\alpha,\beta=-\infty}^{\infty} \vec
b_{\alpha\beta}\exp\left\{-i\left [\vec \kappa_
{\alpha\beta}\vec \rho - \gamma(\vec \kappa_{\alpha\beta})(z+h)  \right ] \right\} ,~z \le -h,  \\
\end{aligned}
\end{equation}
where $\gamma(\vec
\kappa_{\alpha\beta})=\sqrt{k^2-\kappa^2_{\alpha\beta}}$, $\vec
\kappa_{\alpha\beta}=\vec e_x 2\pi\alpha/d +\vec e_y 2\pi\beta/d$,
and $\vec a_{\alpha\beta} =\left\{ \vec E_{1t}(\vec J_1, \vec J_2)
\right\}(\vec \kappa_{\alpha\beta})/d^{2}$, $\vec b_{\alpha\beta} =
\left\{\vec E_{2t}(\vec J_1, \vec J_2) \right\}(\vec
\kappa_{\alpha\beta})/d^{2}$.

If the period of the gratings is less than the wavelength $d/\lambda
< 1$, then in the reflected and transmitted fields in the far-field
zone there exists only the fundamental spatial harmonics ($\vec
a_{00}$, $\vec b_{00}$), which propagate along the normal to the
structure (single-wave mode). On the other hand, at the chosen
polarization of the incident field, the studied structure does not
provide any polarization conversion of the scattered field. So, the
magnitudes $\vec a_{00}$ and $\vec b_{00}$ become to be scalar
quantities, and the reflection and transmission coefficients can be
defined as $R=a_{00}/A_0$ and $T=b_{00}/A_0$. Consequently, in this
paper, the analysis of the diffraction problem is carried out in
this single-wave mode (for extra details the reader is referred to
Ref. \cite{mladyonov_PhysicsAstronomy_2010}).

Thus, the obtained solution allows us to calculate the magnitude and
distribution of the current $J$ along the strips, the reflection $R$
and transmission $T$ coefficients as functions of normalized
frequency ($\ae=d/\lambda$), permittivity ($\varepsilon$) and other
parameters of the structure
\begin{equation}\label{eq:linear}
J=J(\ae,\varepsilon),\quad R=R(\ae, \varepsilon),\quad
T=T(\ae,\varepsilon).
\end{equation}
Noteworthy that the obtained solution has a good experimental
confirmation \cite{fedotov_PhysRevE_2005}.

\section{Spectral properties and bistability \label{sec:spectral}}

\subsection{Trapped-mode resonances}\label{sec:linear}

Trapped-mode resonances can be excited in planar metamaterials with
particles of different shape in the case when these particles
possess specific structural asymmetry. These resonances are the
result of antiphase current oscillations with almost the same
magnitudes which appear on the particles parts (arcs). The scattered
electromagnetic far-field produced by such current oscillations is
very weak, which drastically reduces the field coupling to free
space and therefore radiation losses. As a consequence, the strength
of the induced current can reach very high value and therefore
ensure a high-$Q$ resonant optical response. Such a resonant regime
is referred to so-called ``trapped-modes".

\begin{figure}[htb]
\centerline{\includegraphics[width=16.0cm]{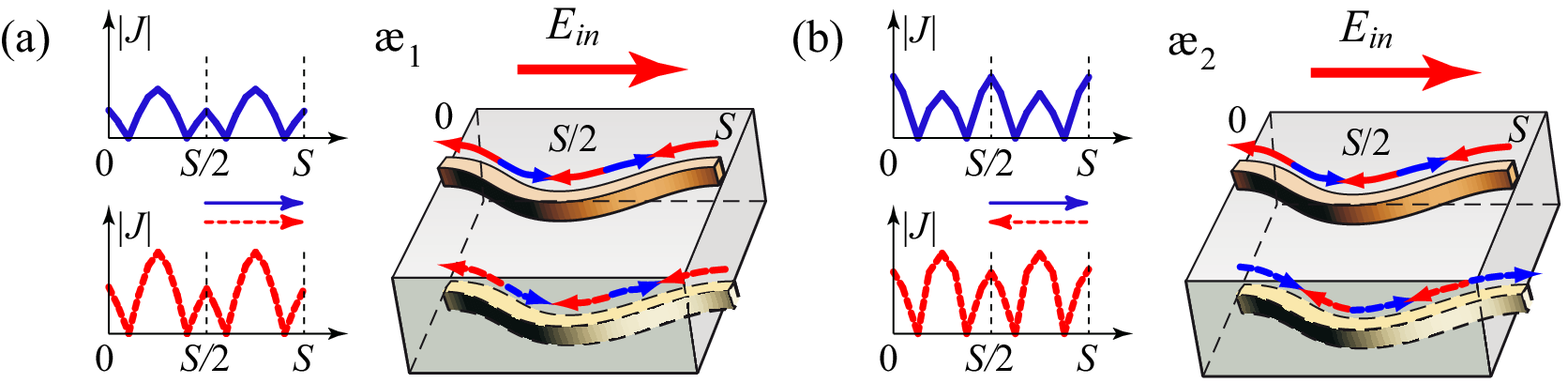}} \caption{}
\label{fig:fig2}
\end{figure}

Due to the bilayer configuration of the structure under study there
are two possible current distributions which correspond to the
trapped-mode resonances. The first distribution is the antiphase
current oscillations in arcs of each grating. In this case the
structure can be considered as a system of two coupled resonators
which operate at the same frequency because the gratings are
identical. We have labeled this resonant frequency in
Figs.~\ref{fig:fig2},~\ref{fig:fig3} by the letter $\ae_1$.
Obviously the distance $h$ between the gratings strongly determines
the resonant frequency position since this parameter defines the
electromagnetic coupling strength. The $Q$-factor of this resonance
is higher in the bilayer structure than that one existed in a
single-layer structure \cite{prosvirnin_JElectromagWavesAppl_2002}
but their similarity lies in the fact that the current magnitude in
the metallic pattern depends relatively weakly on the thickness and
permittivity of the substrate.

The second distribution is the antiphase current oscillations
excited between two adjacent gratings. Similarly we have labeled
this resonant frequency in Figs.~\ref{fig:fig2},~\ref{fig:fig3} by
the letter $\ae_2$. As is known, the closer are the interacting
metallic elements, the higher is the $Q$-factor of the trapped-mode
resonance. Thus varying both the distance between the gratings and
substrate permittivity changes the trapped-mode resonant conditions
and this changing manifests itself in the current magnitude.
Remarkably, in the situation of such current distribution, the field
is localized between the gratings, i.e. directly in the substrate,
which can sufficiently enhance the nonlinear effects if the slab is
made of a field intensity dependent material.

\subsection{Nonlinear spectral properties}\label{sec:nonlinear}

Next we suppose that the structure substrate is a Kerr nonlinear
dielectric which permittivity $\varepsilon$ depends on the intensity
of the electromagnetic field inside it
($\varepsilon=\varepsilon_1+\varepsilon_2|E_{in}|^2$). In this
study, we suppose that operations are provided in the infrared part
of the spectrum where the trapped-mode resonances in metamaterials
with metallic pattern are still well observed and have a high
$Q$-factor \cite{khardikov_JOpt_2010}. In this range the nonlinear
slab can be made of some semiconductor material (e.g. GaAs or InSb).
Based on \cite{palik}, we expect that nonlinear response of the
proposed structure becomes apparent at the input light intensity
about 1-10~kW~cm$^{-2}$. Nevertheless, the effects discussed in the
article can be actual to a broader range of structures including
systems in the microwave range, where nonlinearity is brought about
by lumped elements (e.g. transistor-loaded metamaterials).

Under rigorous consideration, the nonlinear permittivity
$\varepsilon$ of the substrate is inhomogeneous. The permittivity
riches its maximum value directly under the metallic pattern and
along the strips this permittivity is also different. Nevertheless
at the trapped-mode resonance, the flow of electromagnetic energy is
confined to a very small region between the strips and the crucial
impact of the permittivity on the system features occurs in this
place. Therefore, it is possible to construct an approximate
treatment to estimate the inner field intensity and to solve the
\textit{nonlinear problem}. This approximate treatment is based on
two points. The first one takes into account the transmission line
theory and postulates that the inner field intensity is directly
proportional to the square of the current magnitude averaged over a
metal pattern extent \cite{prosvirnin_JElectromagWavesAppl_2002,
tuz_EurPhysJApplPhys_2011}. The second approximation is related to
the homogenization procedure which is convenient to the metamaterial
methodology \cite{poutrina_NewJPhys_2010}. It assumes that, in view
of the smallness of the elementary translation cell of the array
($d<\lambda$), the nonlinear substrate remains to be a homogeneous
dielectric slab under an action of the intensive light.

In more details, according to this theory, the reference meander
wire is considered as a conducting wire periodically loaded by
short-circuited sections of transmission lines (each section
represents one period). Along these wires the currents flow in
opposite directions. Thus the inner electric field strength is
defined as $E^{in} = V/D$, where $V = Z \bar J$ is the line voltage,
$D$ is the distance between wires in the transmission line, $\bar J$
is the current magnitude averaged along the wire. The impedance is
determined at the corresponding resonant frequency $\ae_j =
d/\lambda_j$, $j=1,2$. So, it is $Z=i(Z_0/\pi
d)\cosh^{-1}(D/2\tau_0)\tan(k\Delta)$, where $Z_0$ is the impedance
of free space, and $D=d/2$ and $D=h$ at the the frequencies $\ae_1$
and $\ae_2$, respectively; $\tau_0$ is the wire radius, and $\Delta$
is the length of the equivalent line section. For the equivalent
wire radius there is a simple estimate $\tau_0=w/2$
\cite{Yatsenko_2003}.

From this model it follows that the electric field strength between
the wires is directly proportional to the average current magnitude
$\bar J$ ($I_{in}=|E^{in}|^2\sim\Bar J^2$). Thus, the nonlinear
equation on the average current magnitude in the metallic pattern
can be obtained in the form \cite{tuz_PhysRevB_2010,
tuz_EurPhysJApplPhys_2011, tuz_JOpt_2012}
\begin{equation}\label{eq:nonlinear}
\Bar J=\tilde A F_{\Bar J}(\ae,
\varepsilon_1+\varepsilon_2(I_{in}(\Bar J))),
\end{equation}
where $\tilde A$ is a dimensionless coefficient which depicts how
many times the incident field magnitude $A_0$ is greater than 1 V
cm$^{-1}$. The input field magnitude $A_0$ is a parameter of this
nonlinear equation. So, at a fixed frequency $\ae$, the solution of
this equation gives us the averaged current magnitude $\bar J$ which
depends on the magnitude of the incident field $A_0$.

When implementing our computational algorithm, the approach
described in Section~\ref{sec:structure} is included as a part
inside the subroutine of the nonlinear equation
(\ref{eq:nonlinear}). The input parameters of this subroutine are
the frequency $\ae$, magnitude $A_0$ and inner field intensity
$I_{in}$. The output parameter is the residual between the inner
field intensity passed to the subroutine and the one calculated
inside the subroutine body. The subroutine is called iteratively
until the resulting residual becomes less than a predetermined
threshold. Note that such a solution may give more than one result,
which is a distinctive feature of the nonlinear equation.

Thus, on the basis of the current $\bar J(A_0)$ found by a numerical
solution of the nonlinear equation, the actual value of permittivity
$\varepsilon$ of the nonlinear slab is determined and the reflection
$R$ and transmission $T$ coefficients are calculated as functions of
the frequency $\ae$ and the magnitude of the incident field $A_0$
\begin{equation}\label{eq:coeff}
R=R(\ae, \varepsilon_1+\varepsilon_2(I_{in}(A_0))),\quad T=T(\ae,
\varepsilon_1+\varepsilon_2(I_{in}(A_0))).
\end{equation}

One can see that as the magnitude of the incident field rises, the
frequency dependences of the inner intensity acquire a form of bent
resonances (Fig.~\ref{fig:fig4}) and so-called bistable regime
occurs. The nature of this effect is studied quite well
\cite{gibbs_book_1985,tuz_PhysRevB_2010, tuz_EurPhysJApplPhys_2011,
tuz_JOpt_2012}. Among others, the resonant frequencies of a system
are crucially determined by the permittivity of the material of
substrate. Thus, when the frequency of the incident wave is tuned
nearly the resonant frequency, the field localization produces
growing the inner light intensity which can alter the permittivity
enough to shift the resonant frequency. When this shift brings the
excitation closer to the resonant condition, even more field is
localized in the system, which further enhances the shift of
resonance. This positive feedback leads to formation of the
hysteresis loop in the inner field intensity with respect to the
incident field magnitude, and, as a result, under a certain
magnitude of the incident field, the frequency dependences of the
inner field intensity take a form of bent resonances.

An important point is that this bending is different for distinctive
resonances due to difference in their current magnitudes. It results
in a specific distortion of the curves of the transmission
coefficient magnitude nearly the trapped-mode resonant frequencies.
Thus, at the frequency $\ae_1\sim0.78$ the inner field produced by
antiphase current oscillations is confined in the area in the
vicinity of each grating and it weakly affects on the permittivity
of dielectric substrate. In this case the resonant curve acquires a
transformation into a closed loop which is a dedicated
characteristic of sharp nonlinear Fano-shaped resonances
\cite{miroshnichenko_RevModPhys_2010}. The second resonance
$\ae_2\sim0.82$ is smooth but the current oscillations produce the
strong field concentration between two adjacent gratings directly
inside the dielectric substrate. It leads to a considerable
distortion of the transmission coefficient magnitude in a wide
frequency range, and at a certain incident field magnitude this
resonance reaches the first one and tends to overlap it.
(Fig.~\ref{fig:fig4}b). Evidently, in this case, the transmission
coefficient acquires more than two stable states, i.e. the effect of
multistability arises.

\section{Time-domain simulation and all-optical switching\label{sec:timedomain}}
\subsection{Numerical technique description\label{sec:numdescription}}

The existence of bistability is a necessary but by no means a
sufficient condition for practical realization of bistable
switching, since the conditions needed to excite a particular branch
of the hysteresis loop remain to be determined. Moreover, a
hysteresis in the frequency is of limited use because a continuous
sweep of the frequency would often be impractical in communication
applications. On the other hand modeling in the frequency domain
gives no information on dynamics of response and pulse propagation
through a nonlinear metamaterial. Therefore, in this section we
demonstrate bistable switching in a bilayer fish-scale metamaterial
using time-domain simulations.

On the basis of the microscopic Lorentz-theory approach
\cite{PhysRevB.86.075138} the effective dielectric permittivity
$\varepsilon_{eff}$ can be obtained in order to study dynamic
behavior of the proposed structure in the time-domain (see Appendix~
\ref{sec:appendixA})
\begin{eqnarray}
\varepsilon_{eff} (\ae)= \varepsilon+\frac{\alpha_1}{\ae_1^2
\varepsilon_1/\varepsilon-\ae^2-i \gamma_1
\ae}+\frac{\alpha_2}{\ae_2^2 \varepsilon_1/\varepsilon-\ae^2-i
\gamma_2 \ae}, \label{eps}
\end{eqnarray}
where $\varepsilon$ is permittivity of the nonlinear substrate, and
$\varepsilon_1$ is its linear part. Further we assume the problem
parameters in Eq.~\eqref{eps} to be (in the normalized units):
$\ae_1=0.785$, $\ae_2=0.82$, $\alpha_1=0.01$, $\alpha_2=0.075$,
$\gamma_1=0.0008$, $\gamma_2=0.005$, and $\varepsilon_1=3$. The
resulting effective dielectric permittivity of the homogenized layer
and corresponding transmission spectra are shown in Fig.~\ref{fig5}.
One can see that the obtained curves of transmission spectra  are
characterized by two asymmetric Fano-shaped resonances, which
qualitatively agrees with the results calculated rigorously using
the method of moments (see Fig.~\ref{fig:fig3}).

In order to model light propagation through such an effective
medium, we must obtain the polarization corresponding to
$\varepsilon_{eff}$. This polarization can be found if we assume
that both resonances are in accordance with the oscillations of
polarization in the form
\begin{eqnarray}
\frac{d^2P_k}{dt^2}+\gamma_k
\frac{dP_k}{dt}+\frac{\varepsilon_1}{\varepsilon} \omega^2_k
P_k=\alpha_k E, \qquad k=1,2, \label{oscil}
\end{eqnarray}
where $E$ is the electric field strength and $\omega_k$ are the
resonant frequencies. The resulting electric induction is
$D=\varepsilon E + P_1 + P_2$. Note that this approach is analogous
to that one used in procedures of the Meep package
\cite{OskooiRo10}. Assuming $P_k=p_k (t) \exp{(i \omega t)}$ and
$E=A (t) \exp{(i \omega t)}$, equations \eqref{oscil} can be
rewritten for the amplitudes of polarization $p_k$,
\begin{eqnarray}
\frac{d^2p_k}{d\tau^2}+\eta_k \frac{dp_k}{d\tau}+\zeta_k
p_k=\alpha_k A(t), \qquad k=1,2, \label{oscil1}
\end{eqnarray}
where $\tau=\omega t$, $\eta_k=2 i + \gamma_k$, $\zeta_k=\omega^2_k
\varepsilon_1/\varepsilon - 1 + i \gamma_k$, and all the values
$\gamma_k$, $\omega_k$, and $\alpha_k$ are normalized by the light
frequency $\omega$. Equation \eqref{oscil1} can now be easily
discretized, so that the value of polarization at every time instant
$\tau^l=l \Delta \tau$ can be calculated by
\begin{eqnarray}
p^{l+1}_k=[p^l_k (2-\zeta_k \Delta \tau^2) + p^{l-1}_k (-1+\eta_k
\Delta \tau/2) + \alpha_k \Delta \tau^2 A^l]/(1+\eta_k \Delta
\tau/2), \qquad k=1,2. \label{polar}
\end{eqnarray}
It is worth to note that the values $\zeta_k$ contain the nonlinear
dielectric permittivity $\varepsilon = \varepsilon_1 + \varepsilon_2
|A|^2$ and, hence, depend on the strength of the electric field as
well.

The propagation of light wave in the effective medium is governed by
the wave equation
\begin{eqnarray}
\frac{\partial^2 E}{\partial z^2} = \frac{1}{c^2}\frac{\partial^2
D}{\partial t^2}, \label{weq1}
\end{eqnarray}
where, as was mentioned above, $D$ contains the contributions of the
polarizations $P_k$. Equation~\eqref{weq1} can be solved numerically
using the finite-difference time-domain (FDTD) technique. We
represent the field strength as $E=A (t,z) \exp{[i( \omega t-kz)]}$,
where $\omega$ is a carrier frequency, $k=\omega /c$ is the
wavenumber, and then solve equations for the complex magnitude $A
(t,z)$. Finally, introducing dimensionless arguments $\tau=\omega t$
and $\xi=kz$, the scheme of calculation of the magnitude values at
the mesh points $(l\Delta\tau, j\Delta\xi)$ is realized as follows
\begin{eqnarray}
A_j^{l+1}=[-a_1 \varepsilon^{l-1}_j A_j^{l-1} + b_1 A_{j+1}^l + b_2
A_{j-1}^l + (f \varepsilon^{l}_j - g) A_j^l - R^l_j]/a_2
\varepsilon^{l+1}_j. \label{scheme1}
\end{eqnarray}
Here the auxiliary values are $a_1=1-i \Delta\tau$, $a_2=1+i
\Delta\tau$, $b_1=(\Delta\tau/\Delta\xi)^2 (1-i\Delta\xi)$,
$b_2=(\Delta\tau/\Delta\xi)^2 (1+i\Delta\xi)$, $f=2+\Delta\tau^2$,
$g=2(\Delta\tau/\Delta\xi)^2+\Delta\tau^2$, and the term responsible
for resonant polarizations is
\begin{eqnarray}
R_j^l= \sum_{k=1}^2 [a_2 p^{l+1}_j+a_1 p_j^{l-1} - f p_j^l]_k.
\label{scheme1}
\end{eqnarray}

The stability of the algorithm is provided by the standard Courant
condition between the step intervals of the mesh; in our notation it
can be written as $\Delta\tau/\Delta\xi \leq n$, where $n$ is the
refractive index. At the edges of the calculation region, we apply
the so-called absorbing boundary conditions using Total Field /
Scattered Field (TF/SF) and Perfectly Matched Layer (PML) techniques
\cite{Taflove}.

\subsection{Dynamics of response and pulse propagation\label{sec:dynamics}}

The results of calculations using the method described above are
shown in Fig.~\ref{fig6}. The parameters of the medium are the same
as previously, except the thickness which is now taken to be
$h/d=0.5$. The dimensionless frequency of the incident light is
$\ae=0.783$, i.e. it is chosen to be nearly the  frequency of the
first trapped-mode resonance (Fig.~\ref{fig:fig3}). The normalized
magnitude $A/A_0$ of light varies from $1$ to $11$ with a certain
period which is sufficient for steady-state establishment. The
magnitude of normalization $A_0$ is such that nonlinearity
coefficient satisfies condition $n_2 A_0^2=0.001$, where $n_2
\approx \varepsilon_2/2 \sqrt{\varepsilon_1}$. The curves plotted in
Fig.~\ref{fig6} clearly demonstrate bistable behavior of the system:
panels (a) and (b) represent the transmission and reflection
magnitudes as functions of the incident light magnitude. Note that
the rest of the light energy is absorbed by the layer. Also the
panel (c) shows the hysteresis of the stationary value of light
intensity in the middle of the layer.

Recall that the main feature of the considered fish-scale
metamaterial is the appearance of two particular trapped-mode
resonances. The same two resonances can be seen if we plot the
effective permittivity $\varepsilon_{eff}$ as a function of the
nonlinear contribution $\varepsilon$ and, hence, intensity.
Therefore, if the range of magnitude variation is large enough, one
can expect to obtain two hysteresis loops, which is the indication
of multistability. In order to reduce absorption and the necessary
input intensities, we took the thinner layer than in previous
example, $h/d=0.2$, and get closer to the resonance $\ae=0.784$;
other parameters are left the same. The results of calculations are
shown in Fig.~\ref{fig7}. It turned out that the appearance of the
second loop strongly depends on the initial light magnitude.
Starting from low input intensities and increasing it with a certain
step, we obtained the transition between distinct stable states
corresponding to the shift of the first resonance, but we were not
able to observe the similar jump associated with the second
resonance (see Fig.~\ref{fig7}(a)). Perhaps, the transition
magnitude is very high, so that we could not reach it. However, we
can accept the opposite strategy: start just from the high enough
magnitude and decrease it. In this case, we obtain the clear
evidence of the second loop as can be seen in Fig.~\ref{fig7}(b).
The characteristics of the first loop are identical in both cases.

Another way to study bistable response of our nonlinear metamaterial
is to launch a pulse into it and analyze the shape of the resulting
radiation profile. We consider a Gaussian pulse with magnitude
$A=A_p \exp(-t^2/2t_p^2)$, where $t_p=N \lambda/c$ is its duration
measured through the number of periods $N$. In our calculations,
$N=200$ and $A_p=20 A_0$. The dynamics of transmitted and reflected
light are presented in Fig.~\ref{fig8}(a). The characteristic
features of pulse interaction with bistable medium are seen -- the
flattening of the profile and the overshoot at the leading edge of
the pulse \cite{gibbs_book_1985}. The bistable switching is not
evident in these figures and appears through the asymmetric form of
the output pulse. To reconstruct the hysteresis loop, we calculate
the ratio of the transmitted intensity to the input one at every
time instant (the propagation time through the layer is neglected)
and obtain the dependence of the transmission on the input intensity
for both rising and decaying pulse slopes. The bistable curve
derived from the transmitted pulse profile (Fig.~\ref{fig8}(b)) is
in agreement with the results of stationary calculations of
Fig.~\ref{fig7}(a). To prove the existence of the second loop, we
launched not a full pulse, but only a half of it. The profile of
transmitted radiation given in Fig.~\ref{fig8}(c) clearly shows the
occurrence of two different stable states. The corresponding loop is
presented in Fig.~\ref{fig8}(d) where the curve for ascending
intensities is taken from the panel (b). Thus, the results of
stationary and pulsed calculations are in full accordance with each
other.

Finally, we should touch on the problem of modulation instability
(MI) \cite{Zharov_JApplPhys_2009, Lapshina_OptLett_2012,
Lapshina_JETPLett_2013}. Though there is not any evidences of
self-pulsations or other MI effects in the stationary regime
(Figs.~\ref{fig6} and \ref{fig7}) and in the case of the full pulse
(Fig.~\ref{fig8}(a)), the beatings in Fig.~\ref{fig8}(c) can be
considered as a probable manifestation of MI. To make the final
decision, the careful study of the conditions for MI appearance in
the structure considered should be done which we plan to perform in
our future investigations.

\section{Conclusions\label{sec:conclusions}}

The light transmission through a bilayer fish-scale metamaterial is
considered in both frequency and time domains. An important feature
of the studied structure is its ability to support two distinct
configurations of the trapped-mode resonance excitation. The first
configuration corresponds to a specific distribution of currents,
which oscillate in the same manner on each grating. The second
configuration is a result of antiphase oscillations of currents
which flow in opposite direction relative to the adjacent gratings.
In the latter case it is found out that the electromagnetic field is
confined between the gratings, i.e. directly in the substrate which
leads to manifestation of the effects of bistability and
multistability. They appear in the form of a significant distortion
of transmission coefficient magnitude over a wide frequency band. In
addition a study of the pulse propagation through the metamaterial
is investigated. The results of stationary and pulsed calculations
are in full compliance with each other.

We argue that more parameters can be controlled in the studied
metamaterial and it has better performance than in metamaterials
with a unit cell of a single symmetrical resonant element.This
independent control provides an alternative route to the
optimization of nonlinear materials, which form the basis of such
optical devices as mixers, frequency doublers and modulators.

\bigskip

This work was supported by the Ukrainian State Foundation for Basic
Research, project F54.1/004, and the Belarusian Foundation for Basic
Research, project F13K-009.

\appendix
\section{Appendix \label{sec:appendixA}}

An electron in the metal is affected by the external electric field
$E$ as well as by the Coulomb force $F_C$ from the other electric
charges $q_j$ and electromotive force $F_L$ from the electric
currents. The Coulomb force is the restoring force described as in
the case of harmonic oscillator by $F_C = \sum_j \frac{e
q_j}{\varepsilon r_j} = - m \omega_0^2 \xi$, where $e$ and $m$ are
the electron charge and mass, $\xi$ is electron's displacement. It
is important that the force and the oscillator (resonant) frequency
$\omega_0^2$ are inversely proportional to the dielectric
permittivity of the nonlinear substrate $\varepsilon$. In the linear
and nonlinear regimes, we have $\omega_{01}^2 = a/\varepsilon_1$ and
$\omega_{0}^2 = a/\varepsilon$, respectively, where $\varepsilon_1$
is the linear part of $\varepsilon$. Thus, $\omega_{0}^2 =
\omega_{01}^2 \varepsilon_1/\varepsilon$. The electromotive force is
determined by the magnetic field induced by the electric current $j
= e N
\partial \xi/\partial t$ and takes the form $F_L = -e \partial
\Phi/c
\partial t \sim
\partial B/\partial t \sim \partial j/\partial t \sim \partial^2
\xi/\partial t^2$, where $N$ is the density of electrons, $\Phi$ is
the magnetic flux. We present this force by means of the coefficient
$\beta$, so that $F_L = - \beta m
\partial^2 \xi/\partial t^2 = - \beta m \ddot \xi$. Thus, the equation
of motion for the electron is
\begin{equation}
m \ddot \xi + m \gamma \dot \xi = e E - m \omega_0^2 \xi - \beta m
\ddot \xi.
\end{equation}
Implying the stationary time dependence $\exp(i \omega t)$ for the
field and displacement, we get
\begin{equation}
\xi = \frac{eE/m(1+\beta)}{\omega_0^2/(1+\beta) - \omega^2 - i\omega
\gamma/(1+\beta)}.
\end{equation}
Polarization of the metal element is determined by $P_m = e N \xi$,
thus resulting in the dielectric permittivity
\begin{equation}
\varepsilon_m = 1 + \frac{4\pi P_m}{E} = 1 +
\frac{\omega_p^2}{\omega_r^2 - \omega^2 - i\omega \gamma_r},
\end{equation}
where $\omega_p^2 = 4\pi e/m(1+\beta)$ is the plasma frequency,
$\omega_r^2 = \omega_0^2/(1+\beta)$ is the resonant frequency, and
$\gamma_r = \gamma/(1+\beta)$ is the collision (damping) frequency.
As one can see, all these frequencies can be diminished by the
coefficient $\beta$. It should be noted that the resonant frequency
in the linear $\omega_{r1}$ and nonlinear $\omega_r$ regimes are
related through $\omega_r^2 = \omega_{r1}^2
\varepsilon_1/\varepsilon$.

Since the volume of metal $V_m$ is much less than the volume of
dielectric $V_d$ and optical response of the metamaterial is
characterized by two resonant frequencies, we can represent the
effective permittivity of the structure as a sum of three entities:
permittivity of the nonlinear substrate $\varepsilon$ and two
resonant terms. In the dimensionless notation, we can write
\begin{eqnarray}
\varepsilon_{eff} (\ae)= \frac{V_d \varepsilon + V_m
\varepsilon_m}{V_d + V_m} \approx
\varepsilon+\frac{\alpha_1}{\ae_1^2
\varepsilon_1/\varepsilon-\ae^2-i \gamma_1
\ae}+\frac{\alpha_2}{\ae_2^2 \varepsilon_1/\varepsilon-\ae^2-i
\gamma_2 \ae}, \label{appeps}
\end{eqnarray}
where $\ae_1$ and $\ae_2$ are resonant frequencies, $\gamma_1$ and
$\gamma_2$ are damping coefficients, and $\alpha_1$ and $\alpha_2$
are parameters of strength of light-matter interaction (effective
plasma frequencies). The multiplier $\varepsilon_1/\varepsilon$ in
the denominators of the last two terms is responsible for the
spectral shift of resonant frequencies due to nonlinearity.

\newpage

Fig.~1. (Color online) Fragment of a planar bilayer fish-scale
metamaterial and its unit cell.

Fig.~2. (Color online) The surface current distribution along the
strips placed on the upper and bottom sides of the substrate in the
bilayer fish-scale metamaterial.

Fig.~3. (Color online) The frequency dependences of the transmission
coefficient magnitude for different values of the substrate
thickness (linear regime); $\varepsilon = 3$, $2w/d=0.05$,
$\Delta/d=0.25$.

\begin{figure}[htb]
\centerline{\includegraphics[width=11.0cm]{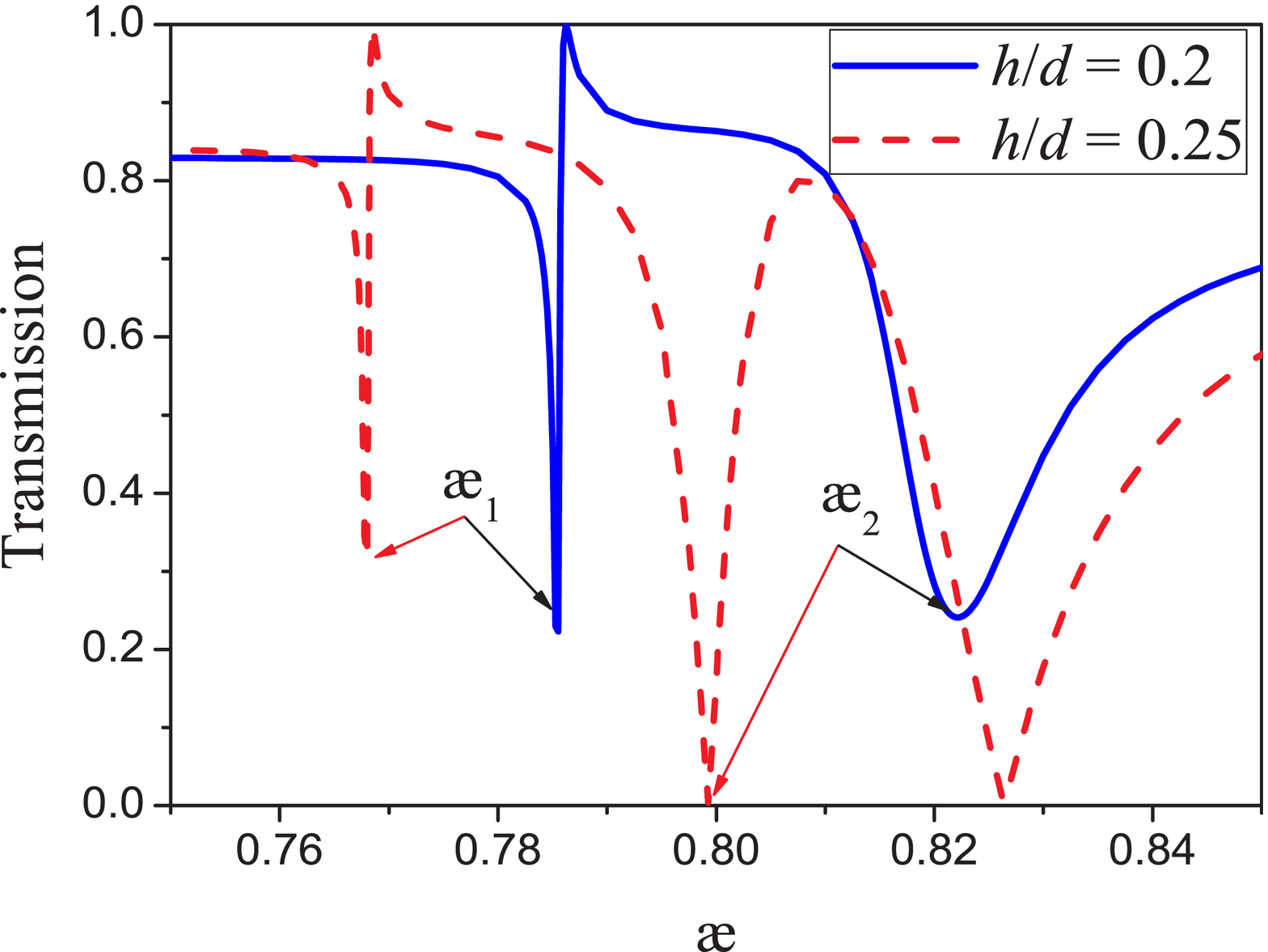}} \caption{}
\label{fig:fig3}
\end{figure}

\begin{figure}[htb]
\centerline{\includegraphics[width=16.0cm]{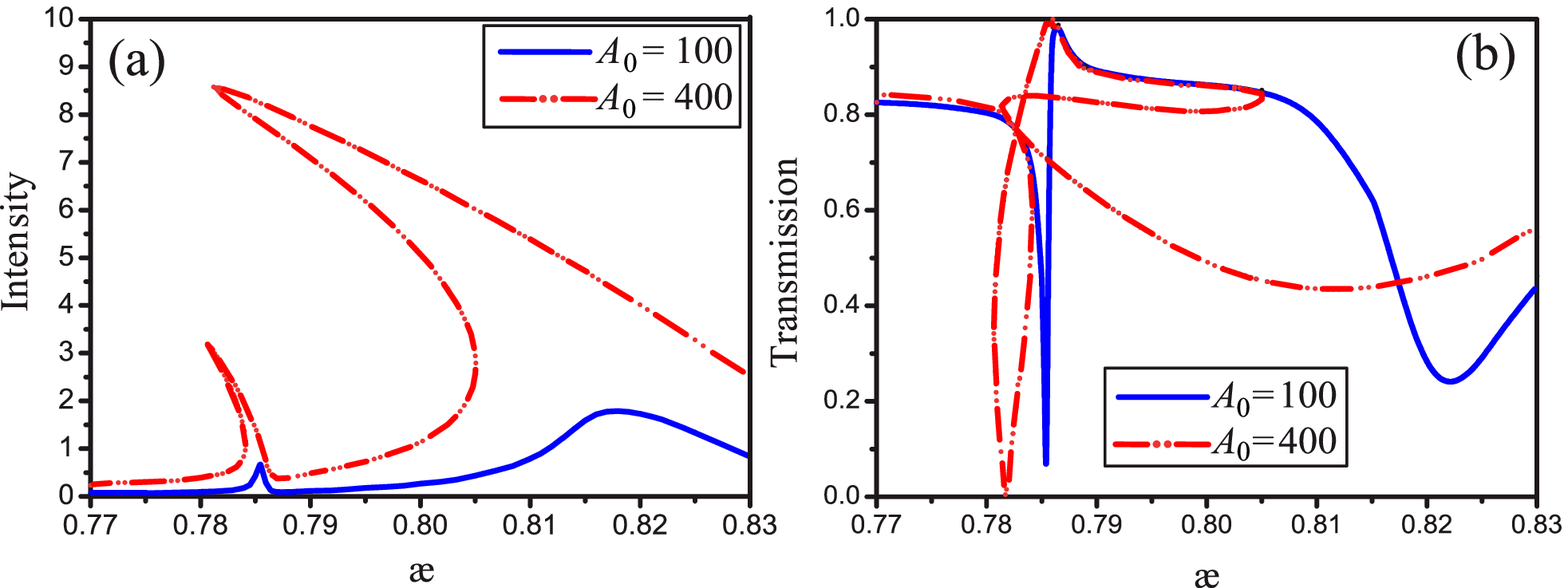}} \caption{}
\label{fig:fig4}
\end{figure}

\begin{figure}[htb]
\includegraphics[width=16.0cm]{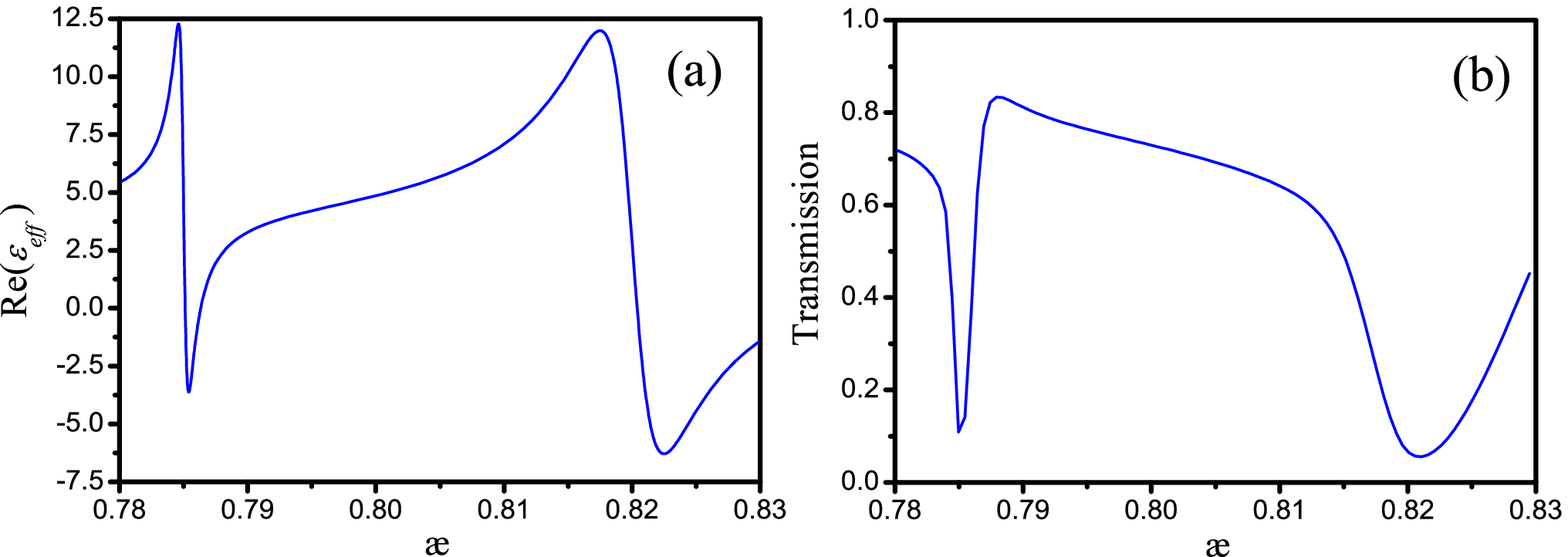}
\caption{} \label{fig5}
\end{figure}

\begin{figure}[htb]
\includegraphics[width=10.0cm]{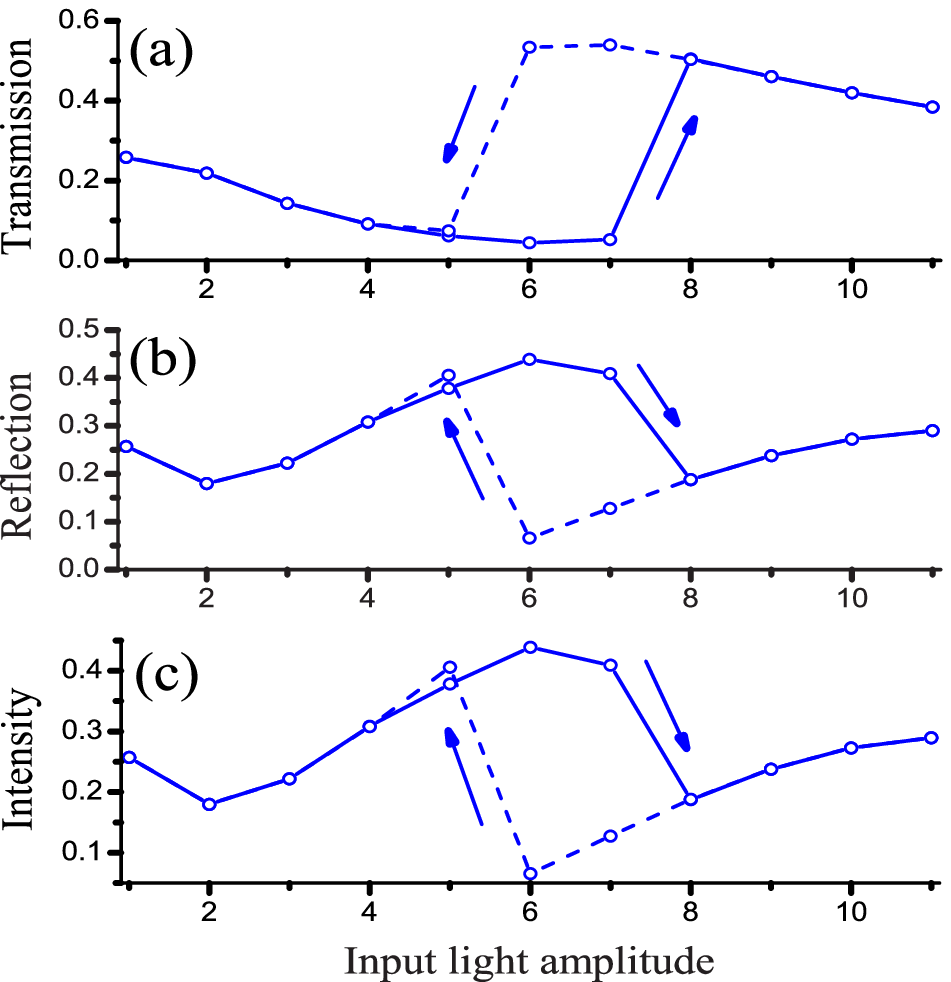}
\caption{} \label{fig6}
\end{figure}

\begin{figure}[htb]
\includegraphics[width=10.0cm]{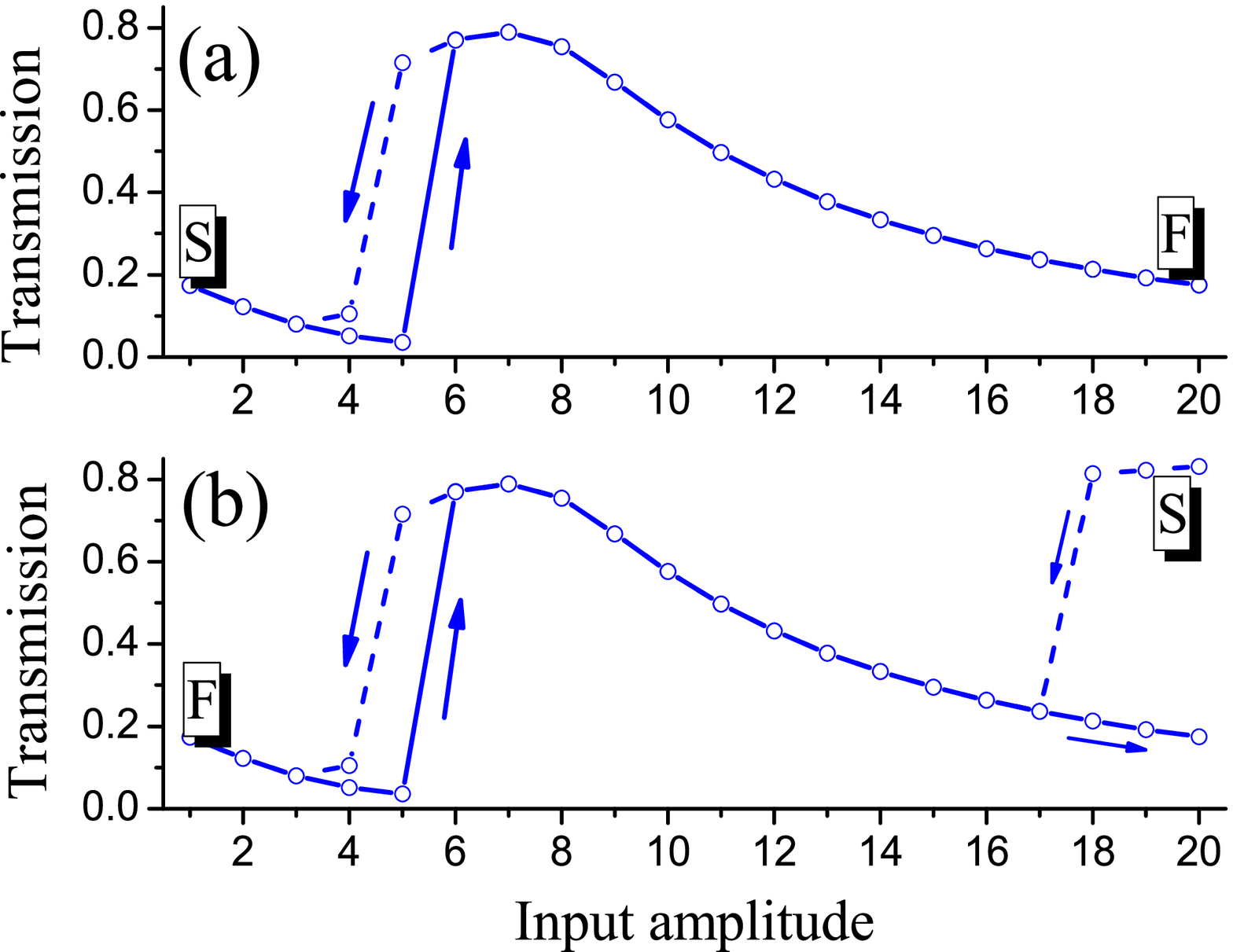}
\caption{} \label{fig7}
\end{figure}

\begin{figure}[htb]
\includegraphics[width=16.0cm]{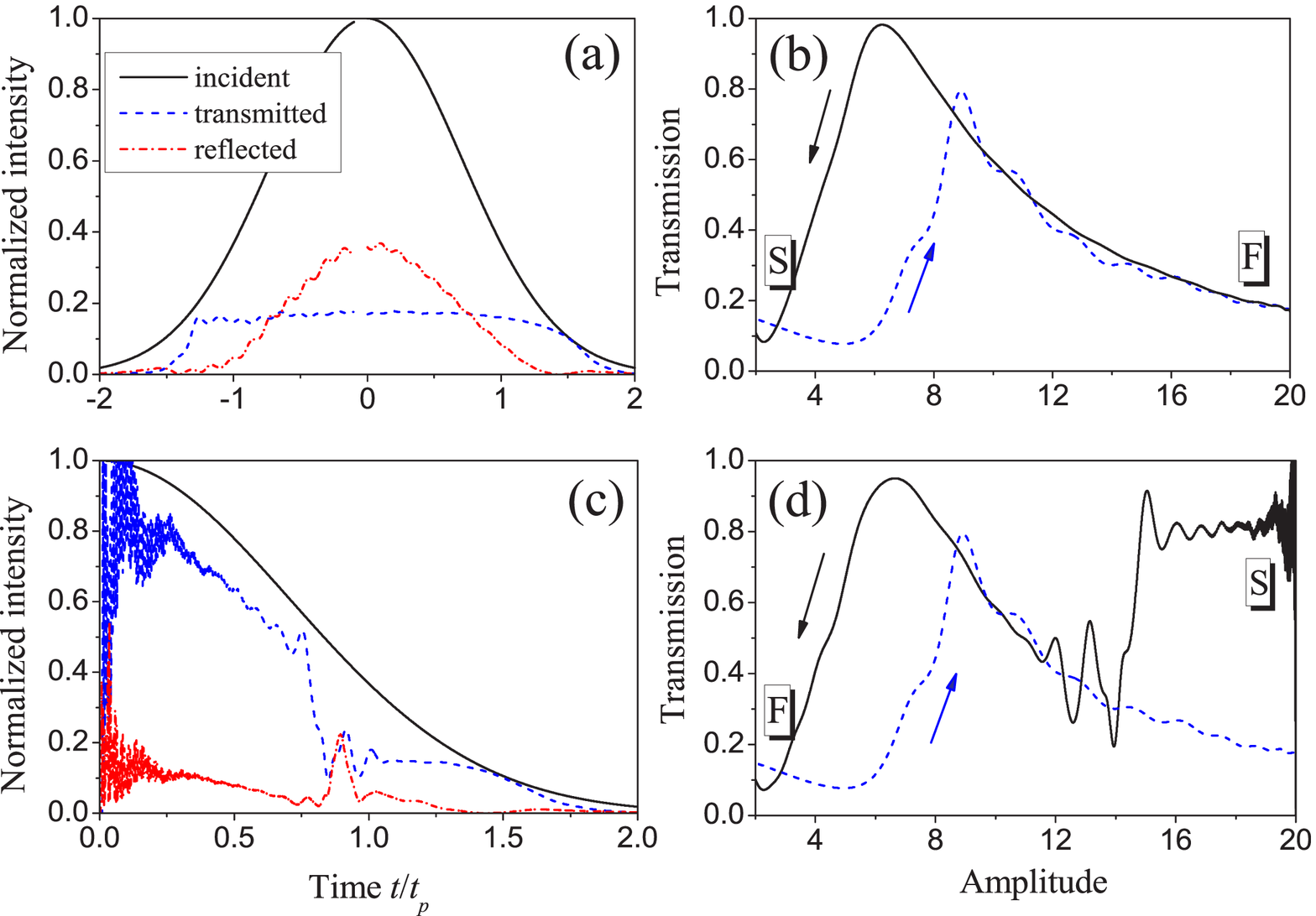}
\caption{} \label{fig8}
\end{figure}

Fig.~4. (Color online) The frequency dependences $(\ae=d/\lambda)$
of (a) the inner field intensity and (b) the transmission
coefficient magnitude  for different values of the incident field
magnitude (nonlinear regime); $h/d=0.2$, $\varepsilon_1=3$,
$\varepsilon_2=0.005$~cm$^2$~kW$^{-1}$. Other parameters are the
same as in Fig.~\ref{fig:fig3}.

Fig.~5. (Color online) (a) Frequency dependency of the effective
dielectric permittivity $\varepsilon_{eff}$ of the homogenized layer
with the thickness $h/d=0.2$ and (b) the corresponding transmission
spectrum.

Fig.~6. (Color online) Bistability curves for stationary levels of
(a) transmission, (b) reflection, (c) normalized intensity in the
middle of the layer.

Fig.~7. (Color online) Bistability curves for stationary levels of
transmission calculated starting from the low magnitude level (a)
and from the high one (b). The letters ``S'' and ``F'' denote the
starting and final values of magnitude variation.

Fig.~8. (Color online) (a) and (c) Profiles of the incident,
transmitted and reflected pulses for the full pulse and the half of
it, respectively. (b) and (d) The bistability curves calculated from
the data of panels (a) and (c). The letters ``S'' and ``F'' denote
the starting and final values of magnitude variation.

\newpage

\end{document}